\documentclass{ifacconf}

\makeatletter \let\cl@part\relax \makeatother 

\usepackage[english]{babel}
\usepackage{xargs}                      
\usepackage{siunitx}

\usepackage{mathematix}
\usepackage[nameinlink]{cleveref}
\usepackage{booktabs}
\usepackage{enumitem}
\usepackage{algorithm}
\usepackage{algpseudocode}
\usepackage{microtype}
\usepackage{units}

\usepackage{nameref}

\definecolor{MainColor}{HTML}{323a54}    
\definecolor{ThemeGreen}{HTML}{61b062}
\definecolor{ThemeRed}{HTML}{FF4422}
\definecolor{AccentColor}{HTML}{3da5d9}%
\definecolor{DarkColor}{HTML}{2364aa}%

\usepackage{pgfplotstable}
\usepackage{filecontents}
\usepackage{graphicx}      
\usepackage{natbib}        
\usepackage{flushend}
\usepackage{accents}

\usepackage{etoolbox}  

\newlist{inlinelist}{enumerate*}{1}
\setlist*[inlinelist,1]{%
  label=(\roman*),
}
\newlist{proofsteps}{enumerate}{1}
\setlist[proofsteps,1]{%
  label=\bfseries{\Roman*},
  itemindent=0pt,
  wide =0.5\parindent,
  listparindent=0pt,%
  afterlabel={{.\nobreakspace}}
}

\crefname{lem}{lemma}{lemmas} 
\crefname{thm}{theorem}{theorems} 
\crefname{def}{definition}{definitions}

\newcommandx{\subscr}[3][3]{
  \ifstrempty{#3}{
    #1_{#2}
  }{
    #1_{#2,#3}
  }
}

\newcommandx{\ego}[2][2]{\subscr{#1}{\scriptscriptstyle{\textsc{EV}}}[#2]}
\newcommandx{\tv}[2][2]{\subscr{#1}{\scriptscriptstyle{\textsc{TV}}}[#2]}
\newcommand{\submax}{\mathrm{max}}
\newcommand{\submin}{\mathrm{min}}

\newcommand{\W}{\mathcal{W}}                
\newcommand{\ve}{\ego{v}}                   
\newcommand{\vt}{\tv{v}}                    
\newcommand{\amax}{a_{\submax}}              
\newcommand{\h}{h}                          
\newcommand{\amin}{a_{\submin}}              
\newcommand{\de}{\ego{\Delta}}                   
\newcommand{\pe}{\ego{p}}            
\newcommand{\pt}{\tv{p}}                   
\newcommand{\at}{\tv{a}}                   

\newcommand{\dt}{\tv{\Delta}}                   
\newcommand{\pvec}{\mu}
\newcommand{\Ts}{T_\mathrm{s}}                     

\newcommand{\rinv}{\mathcal{R}}                  
\newcommand{\maxrinv}{\rinv^{\star}}
\DeclareMathOperator{\pre}{pre}
\newcommand{\Xfeas}{\mathcal{X}}                 
\newcommand{\cmin}{c_{\submin}}
\newcommand{\Xs}{\mathcal{Z}_{\mathrm{s}}}
\newcommand{\tree}{\mathcal{T}}
\newcommand{\ns}{n_x}                            
\newcommand{\row}[2]{#1_{#2}}                   
\newcommand{\elem}[3]{\row{#1}{#2,#3}}           
\newcommand{\idx}[2]{\row{#1}{#2}}
\newcommand{\na}{n_u}                            
\newcommand{\nsample}{n}                         
\newcommand{\hor}{N}                             
\newcommand{\vref}{v_{\mathrm{ref}}}               
\DeclareMathOperator{\TVAVAR}{r-\AVAR}

\newcommand{\cone}{\mathcal{K}}                     
\newcommand{\Ufeas}{\mathcal{U}}

\newcommand{\natseq}[2]{\N_{[#1,#2]}}
\newcommand{\Xr}{\Xfeas_{\mathrm{r}}}      
\newcommand{\cE}[1]{\E_{|#1}}
\newcommand{\w}{\mathbf{w}}                
\newcommand{\Xrss}{\Xfeas_{\scriptscriptstyle{\textsc{RSS}}}}
\newcommand{\nodestree}[2]{\nodes_{#1}(#2)}
\newcommand{\anctree}[2]{\anc_{#1}(#2)}
\newcommand{\childtree}[2]{\child_{#1}(#2)}

\newcommand{\ambSeq}{\bar{\amb}}
\newcommand{\vmax}{v_{\submax}}
\newcommand{\Xc}{\Xfeas_{\textrm{c}}}

\begin{document}
\begin{frontmatter}

\title{Learning-Based Risk-Averse Model Predictive Control for 
Adaptive Cruise Control with Stochastic Driver Models \thanksref{footnoteinfo}} 

\thanks[footnoteinfo]{This work was supported by the Ford-KU Leuven Research Alliance. The work of P. Patrinos was supported by: FWO projects: No. G086318N; No. G086518N; Fonds de la Recherche Scientifique – FNRS, the Fonds Wetenschappelijk Onderzoek – Vlaanderen under EOS Project No. 30468160 (SeLMA), Research Council KU Leuven C1 project No. C14/18/068.}

\author[First]{Mathijs Schuurmans} 
\author[Second]{Alexander Katriniok} 
\author[Third]{Hongtei Eric Tseng}
\author[First]{Panagiotis Patrinos}

\address[First]{Department of Electrical Engineering \textsc{esat-stadius}, KU Leuven, 
Kasteelpark Arenberg 10, 3001 Leuven, Belgium (e-mail: \{mathijs.schuurmans, panos.patrinos\}@esat.kuleuven.be).}
\address[Second]{Ford Research \& Innovation Center,
 52072 Aachen, Germany
   (e-mail: de.alexander.katriniok@ieee.org)}
\address[Third]{Research \& Innovation Center, Ford Research Laboratories, Dearborn, MI 48124 USA}
\begin{abstract}                
    We propose a learning-based, distributionally robust model predictive control approach towards the design of adaptive cruise control (ACC) systems. We model the preceding vehicle as an autonomous stochastic system,
    using a hybrid model with continuous dynamics 
    and discrete, Markovian inputs. We estimate the (unknown) 
    transition probabilities of this model empirically
    using observed mode transitions and simultaneously determine 
    sets of probability vectors (ambiguity sets) around these estimates, that 
    contain the true transition probabilities with high confidence. 
    We then solve a risk-averse optimal control problem that 
    assumes the worst-case distributions in these sets. 
    We furthermore derive a robust terminal constraint set and use it to establish recursive feasibility of the resulting MPC scheme. We validate the theoretical results and demonstrate desirable properties of the scheme through closed-loop simulations.
\end{abstract}


\end{frontmatter}

\section{Introduction}

In recent decades, the usage of adaptive cruise control (ACC) systems
has become widespread in the automotive research and industry, as they have demonstrated numerous benefits 
in terms of safety, fuel efficiency, passenger comfort, etc. (\cite{xiao_comprehensive_2010}).
The term ACC generally refers to longitudinal control 
systems that are aimed at maintaining a user-specified reference velocity, 
while avoiding collisions with preceding vehicles.

To guarantee safety for related automated driving applications, \cite{shalev-shwartz_formal_2017} proposed the \emph{Responsibility-Sensitive Safety} (RSS) framework, which prescribes minimal safety distances for 
ACC systems based on simple vehicle kinematics.
Under natural assumptions on the possible range of acceleration values for the involved vehicles, this safety distance can guarantee collision avoidance in worst-case conditions.
Furthermore, the authors define rules that prescribe how an ACC system should properly respond to violations of this safety distance. Although safe, the prescribed rules are reactive in nature, which may lead to sudden braking maneuvers, reducing passenger comfort and fuel efficiency. 

By contrast, model predictive control (MPC) methods optimize a specified 
performance index based on the predicted evolution of the controlled system
in the near future, which endows the control system with the capability 
to behave proactively, and adapt its actions with respect to potential future events.
However, due to the involvement of human actors, there is an inherent level of uncertainty in the prediction of traffic situations. In order to explicitly account for this uncertainty, stochastic MPC has been a particularly popular approach
(\cite{bichi_stochastic_2010,moser_flexible_2018,mcdonough_stochastic_2013}).

In an attempt to make accurate predictions about the future 
behavior of the lead vehicle, many different driver models have been proposed in the literature (see \cite{wang2014modeling} for a survey). A common approach is to combine continuous physics-based dynamics with a discrete (and potentially stochastic) decision model for the driver (e.g.,~\cite{sadigh2014data,kiencke1999modeling,bichi_stochastic_2010}). In this work, we follow this line of reasoning and model the preceding vehicle using double integrator dynamics, where the driver's inputs are generated 
by a Markov chain. 

A major shortcoming of stochastic MPC approaches is their dependence 
on accurate knowledge of all probability distributions involved in the stochastic model.
Since, in practice, these are estimated based on finitely sized data samples, they may not accurately reflect the true underlying distributions --- we will refer to this uncertainty on probability distributions as 
\emph{ambiguity}. Due to this ambiguity, stochastic controllers may 
perform unreliably with respect to the true distributions.

The main contributions of this paper address these issues in the following manner. First, we generalize the stochastic MPC methodology for ACC systems by adopting a distributionally robust approach, where not only the estimated 
distribution is taken into consideration, but all distributions that 
belong to a so-called \emph{ambiguity set}.
Under the Markovian assumption, we can use concentration inequalities to obtain closed-form expressions for these sets, such that they contain the data-generating distributions with arbitrarily high confidence. 
For the more general case, where this assumption does not hold,
safety of stochastic MPC techniques can still be improved by constructing 
suitable ambiguity sets using statistical techniques such as bootstrapping 
or cross-validation.  

Secondly, we derive a robust control invariant set which can be used as a terminal constraint set in the proposed control formulation, allowing for  guaranteed recursive feasibility of the resulting MPC scheme.
The underlying philosophy for the methodology is to rely on our knowledge of the physical dynamics to guarantee the required level of safety. All available data is then utilized to reduce costs insofar this does not compromise these guarantees.

\subsection{Notation and preliminaries}

Given two integers $a \leq b$, let $\natseq{a}{b} \dfn \{ n \in \N \mid a \leq n \leq b \}$.  
We define the operator $[\argdot]_+$ as $\max\{0, \argdot\}$, where the $\max$ is interpreted element-wise. 
We denote the element of a matrix $P$ at row $i$ and column $j$ as $\elem{P}{i}{j}$ and the $i$th row of a matrix $P$ as $\row{P}{i}$. Similarly, the $i$th element of a vector $x$ is denoted $\idx{x}{i}$. We denote the vector in $\Re^k$ with all elements one as $\1_k \dfn (1)_{i=1}^{k}$. Finally, we define the indicator function as $1_{x=y} \dfn 1$ if $x=y$ and $0$ otherwise.    

\subsubsection{Risk measures and ambiguity} \label{sec:risk-measures}
Let $\Omega$ denote a discrete sample space endowed with the $\sigma$-algebra $\F = 2^{\Omega}$ and probability measure $\prob$, defining the probability space $(\Omega, \F, \prob)$.
For a given random variable $Z : \Omega \rightarrow \Re$, we can collect the possible outcomes of $Z$ in a \emph{random vector} $\Re^{|\Omega|} \ni z = (Z(i))_{i \in \Omega}$. Similarly, a \emph{probability vector} can be defined as $\simplex_{|\Omega|} \ni \pvec = (\idx{\pvec}{i})_{i \in \Omega} = (\prob[\{\omega\}])_{\omega \in \Omega}$, where $\simplex_k \dfn \{ p \in \Re^{k} | \trans{\1_k} p = 1, p \geq 0 \}$ denotes the probability simplex of dimension $k$. A risk measure $\rho : \Re^{|\Omega|} \rightarrow \Re$ is a mapping from the space of possible outcomes of $Z$ to the real line, which we may use to deterministically compare random variables before their outcome is revealed. In particular, we are interested in so-called \emph{coherent} risk measures,
for which the following dual representation exists~\cite[Thm 6.5]{shapiro2009lectures}
\begin{equation} \label{eq:risk-dual}
    \rho[z] = \max_{\mu \in \amb} \E^{\mu}[z].
\end{equation}
Here, $\amb \subseteq \simplex_{|\Omega|}$ is some closed, convex subset 
of the probability simplex, commonly referred to as the \emph{ambiguity 
set} of $\rho$. This dual representation allows for a distributionally 
robust interpretation where, based on a set of data drawn from an unknown 
distribution, the ambiguity set is typically constructed such that it 
contains all probability distributions that are in some sense consistent 
with the data. We will use this perspective explicitly when constructing 
a data-driven MPC scheme in~\Cref{sec:dist-rob-mpc}. For a given ambiguity 
set $\amb$, we will denote the induced risk measure by $\rho^{\amb}$.
We finally remark that the concept of a risk measure can be extended in a 
straightforward manner to \emph{conditional risk mappings} by replacing 
the expectation in~\eqref{eq:risk-dual} with a conditional expectation.


\section{Nominal stochastic MPC}

In this section, we construct a model for the ACC system and formulate a nominal control problem for the simplified case where all involved probability distributions are known.
We use this setting to derive a terminal constraint set that allows us to ensure recursive feasibility of the MPC scheme. In~\Cref{sec:dist-rob-mpc}, we will extend these results to the setting in which all distributions are to be estimated from data.

\subsection{Modeling and problem statement} \label{sec:modeling}

Throughout this paper, we will assume that
the behavior of the vehicle pair can be modelled as a discrete-time Markov jump linear system (MJLS) (\cite{costa2006discrete}), which has dynamics of the form  
\begin{equation} \label{eq:dynamics}
    x_{t+1} = f(x_t, u_t, w_{t+1}) = A(w_{t+1}) x_t + B(w_{t+1}) u_t + p(w_{t+1}),     
\end{equation}
where $x_t \in \Re^{\ns}$ is the state vector $u_t \in \Re^{\na}$ is the input and $\w \dfn (w_t)_{t\in \N}$ is a Markov chain on $(\Omega, \mathcal{F}, \prob)$ with state space $\mathcal{W} \dfn \natseq{1}{\dimProbSpace}$ and transition matrix $\transmat\in \Re^{\dimProbSpace \times \dimProbSpace}$, where  $\transmat_{i,j} = \prob[w_{t}=j \mid w_{t-1}=i]$.
We assume that at any time $t$, both $x_t$ and $w_t$ are observable.

The goal is to select a state feedback law $\kappa: \Re^{\ns} \times \W \rightarrow \Re^{\na}$, such that for all 
$t \in \N$,
\( 
    \kappa(x_t, w_t) \in \Ufeas,
\)
and that for the closed-loop system 
\(
    x_{t+1} = f(x_t, \kappa(x_t, w_t), w_{t+1}) 
\),
the state satisfies  
\begin{subequations} \label{eq:state-constraints-general}
    \begin{align}
    x_t &\in \Xr, \label{eq:constraints-general-robust}\\ 
    \prob[x_{t+1} &\in \Xc \mid x_t, w_{t}] \geq 1 - \delta, \label{eq:constraints-general-chance}
    \end{align}
\end{subequations}
almost surely ($\as$), i.e., for all $(w_i)_{i=0}^{t} \in \W^{t+1}$ such that $\elem{\transmat}{w_t}{i}>0$. 
Here, the set $\Ufeas$, $\Xr$ and $\Xc$ correspond respectively to the 
input constraints, hard state constraints, and soft (probabilistic) state constraints,
which are specified below.   
\begin{figure}[ht!]
    \centering
    \includegraphics{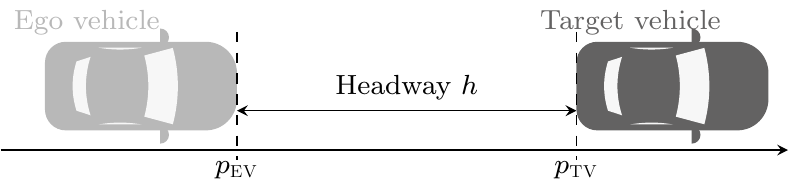}
    \caption{Illustration of the ACC problem.}
    \label{fig:ACC-setup}
\end{figure}
\subsubsection{Dynamics}
We model the longitudinal dynamics of the two vehicles 
along a road-aligned coordinate system and combine 
the states of the ego vehicle and the target vehicle into one system. We denote by $\pe$ and $\pt$ the positions of the ego vehicle and the target vehicle respectively and define $\h \dfn \pt - \pe$ to be the (positive) headway between the two vehicles (see~\Cref{fig:ACC-setup}).
Similarly, we denote the velocities of the ego and target vehicle by $\ve$ and $\vt$, so that the total state of the vehicle pair is described by a state vector 
$x = \trans{\smallmat{\h& \ve& \vt}}$. 

For simplicity, we take the individual vehicle dynamics to be
described by discrete double integrators, such that the combined 
dynamics is given by
\begin{equation} \label{eq:double-integrator-model}
    x_{t+1} = \smallmat{1 & -\Ts & \Ts\\ 0 & 1 & 0\\ 0 & 0 & 1} x_t + \smallmat{0\\\Ts\\0} u_t + \smallmat{0 \\ 0 \\ \Ts \at(x_t, u_t, w_{t+1})},
\end{equation}
where $\Ts$ is the sampling period and $\at$ denotes a mode-dependent acceleration of the target vehicle. Provided that $\at$ is an affine function of the 
states and inputs, this model is compatible with the form~\eqref{eq:dynamics}.
In the remainder of this paper, we assume a parametrization of $\at$ such that in decelerating modes, the input (the brake) behaves like a dissipative element, i.e.,
\begin{equation} \label{eq:TV-model}
    \begin{aligned}
   \at(x,u,w) = \at(x,w) = \begin{cases} 
    c_w,    &\text{if } c_w \geq 0\\
    c_w \idx{x}{3}, &\text{otherwise,} 
   \end{cases}
\end{aligned}
\end{equation}
where $c_w\geq -\nicefrac{1}{\Ts}, w \in \W$ are design parameters. 

\subsubsection{Constraints}
We assume that velocities of the ego vehicle must remain nonnegative and upper bounded by some physical limit $\vmax > 0$, and that the acceleration of the target vehicle is limited between the values 
$\amin \leq 0$ and $\amax \geq 0$. 
This yields the constraint sets 
\begin{subequations} \label{eq:robust-constraint-sets}
\begin{align}
    \Xr &\dfn \{x \in \Re^{\ns} \mid 0 \leq \idx{x}{2} \leq \vmax \}, \label{robust-constraint-set-states} \\ 
    \Ufeas &\dfn \{u \in \Re^{\na} \mid \amin \leq u \leq \amax\} \label{robust-constraint-set-inputs},
\end{align}
\end{subequations}
for the states and inputs respectively. Note that since we assume that the controller has no agency over the target vehicle, we do not pose explicit constraints on the state $x_3$.

Since a stochastic model of the target vehicle will typically include extreme behaviors, albeit with exceedingly small probabilities,
imposing certain safety constraints robustly (i.e., for all possible
realizations of $\w$) will typically lead to overly 
large safety distances, excessive emergency maneuvers,
or even infeasibility of the optimization problem in practically benign situations. It is therefore common to instead impose (conditional) chance constraints of the form~\eqref{eq:constraints-general-chance} (e.g.,~\cite{moser_flexible_2018}). In particular, we want to constrain the headway (possibly defined to include some safety distance), to remain positive: 
\begin{equation*}
    \Xc = \{x \in \Re^{\ns} \mid g(x) = -\idx{x}{1} \leq 0 \}.
\end{equation*}
Since chance constraints~\eqref{eq:constraints-general-chance} are generally nonconvex, it is common to approximate them using risk measures (\cite{nemirovski2012safe}). In particular,
it can be shown~\cite[sec. 6.2.4]{shapiro2009lectures} that for any random variable $z \sim p \in \simplex_{m}$, the following implication holds tightly
\begin{equation} \label{eq:risk-constraint-implication} 
    \AVAR^p_{\delta}[z] \leq 0 \Rightarrow \prob[z \leq 0] \geq 1-\delta.
\end{equation}
Here, $\AVAR^{p}_{\delta}[z]$ denotes a particular risk measure referred to as the \emph{average value-at-risk} (at level $\delta \in (0, 1]$ and with reference probability $p \in \simplex_m$).
It can be defined as~\cite[Thm. 6.2]{shapiro2009lectures}
\begin{equation} \label{eq:def-AVAR}
    \AVAR_{\delta}^p[z] \dfn \E[z \mid z \geq q_{\delta}(z)], 
\end{equation}
where $q_{\delta}(z) \dfn \inf\{t: \prob[z \leq t] \geq 1 - \delta\}$ denotes the $(1-\delta)$-quantile of $z$.
It can furthermore be written in the dual form~\eqref{eq:risk-dual}, with the polytopic ambiguity set  
\begin{equation} \label{eq:ambiguity-avar}
    \amb = \amb_{\AVAR_{\delta}^p} \dfn \{ \mu \in \Re^{|\Omega|} \mid \trans{\1_{|\Omega|}}\mu = 1, 0 \leq \mu \leq \tfrac{p}{\delta} \}.
\end{equation}
By exploiting the structure of ambiguity sets such as $\amb_{\AVAR_{\delta}^{p}}$, \cite{sopasakis2019riskC} show that constraints involving the average value-at-risk can be imposed efficiently using only linear (in)equalities.  
In practice, we can thus satisfy the chance constraint~\eqref{eq:constraints-general-chance} by imposing for $t \in \N$,
\begin{align}
\AVAR_{\delta}^{\transmat_{w_t}}[g(x_{t+1}) \mid x_t, w_t] \leq 0,\; \as \label{eq:constraints-risk}
\end{align}
Note that by virtue of the interpretation \eqref{eq:def-AVAR}, the risk constraints~\eqref{eq:constraints-risk}, additionally to their computational advantages, provide the guarantee of bounding the magnitude of the average chance constraint violation, given that it occurs.

Finally, in order to guarantee recursive feasibility, we impose the final state to be in a robust control invariant set $x_\hor \in \Xfeas_\hor$ for all $(w_i \in \W)_{i \in \natseq{0}{\hor}}$. This set is specified in~\Cref{sec:recursive-feasibility}. 

\subsubsection{Cost function}
We define a stage cost $\ell: \Re^{\ns} \times \Re^{\na}\rightarrow \Re_+$ and terminal cost $\ell_\hor: \Re^{\ns} \rightarrow \Re_+$, that simply assign a quadratic penalty to the deviation from the reference velocity $\vref$ and to the control effort $u$: 
    \begin{align*}
        \ell(x,u) &\dfn q (\idx{x}{2}- \vref)^2 + r u^2, \\
        \ell_\hor(x) &\dfn q (\idx{x}{2}- \vref)^2.
    \end{align*}



\begin{defn}[Nominal stochastic MPC] \label{def:nominal-mpc}
    \sloppy For a given $x \in \Re^{\ns}, w \in \W$,
    the nominal optimal control problem (OCP) comprises of computing an 
    $\hor$-step sequence of admissible policies, i.e., a sequence of functions
    $\pi = (\kappa_i)_{i \in \natseq{0}{\hor-1}}$, with 
    $\kappa_k : \Re^{\ns} \times \W \rightarrow \Re^{\na}$ that 
    solve the optimization problem
        \begin{subequations}\label{eq:nominal-problem}
            \begin{align} \label{eq:nominal-cost}
                \minimize_{u_0}\;& \ell(x_0, u_0) + \inf_{u_1} \cE{0} \Big[ \ell(x_1, u_1)\notag {}+{}\ldots \\
                + \inf_{u_{\hor-1}}& \cE{\hor-2} \big[ \ell(x_{\hor-1}, u_{\hor-1}){}+{} \cE{\hor-1} \big[ \ell_{\hor}(x_\hor) \big] \cdots
                \Big]
                \bigg] 
            \end{align}
            subject to
            \begin{align}
                &x_0= x, w_0 = w, \label{eq:nominal-constraints-initial} \\
                &x_{k+1} = f(x_k, u_k, w_{k+1}) ,\, k \in \natseq{0}{\hor-1},\\
                \label{eq:nominal-MPC-state-input-constraint}
                &u_k = \kappa_k(x_k, w_k) \in \Ufeas, x_k \in \Xr\,\as,\, k \in \natseq{0}{\hor-1},\\ 
                \label{eq:nominal-MPC-crm-constraint}
                &\AVAR_{\delta}^{\transmat_{w_k}}[g(x_{k+1}) \mid x_k, w_k] \leq 0\; \as, k \in \natseq{0}{\hor-1},\\ 
                &x_\hor \in \Xfeas_\hor, \, \as,
            \end{align}
            \end{subequations}
    where $\cE{t}[\argdot] = \E[\argdot | x_t, w_t] $ denotes the conditional expectation given the realization of $(w_{i})_{i=0}^t$.

    The corresponding MPC scheme is obtained by applying the first policy $\kappa_0$ to the system at the current state, and resolving the OCP~\eqref{eq:nominal-problem} in a receding horizon manner.
\end{defn}

\begin{rem}
Note that by linearity of the expectation operator, the cost~\eqref{eq:nominal-cost} is equivalent to the total expectation of the sum of the 
state costs $\ell(x_t, u_t)$ and the terminal cost $\ell_\hor(x_\hor)$. 
However, by writing the cost in the nested form above, we emphasize the 
relation with the risk-averse OCP formulated in~\Cref{sec:dist-rob-mpc}. 
\end{rem}

Due to the discrete nature of $\W$, problem~\eqref{eq:nominal-problem} can be stated as a finite-dimensional optimization problem over a so-called \emph{scenario tree}~(\cite{sopasakis2019risk}).  
A scenario tree $\tree$ (of horizon $\hor$) represents the set of all 
possible realizations of a random process $(w_t)_{t \in \natseq{0}
{\hor}}$, given an initial value $w_0$. We denote 
the set of nodes at stage $t$ of $\tree$ by $\nodestree{t}{\tree}$ so that 
$\{ w^i\}_{i \in \nodestree{t}{\tree}}$ corresponds to all possible 
outcomes of $(w_k)_{k\in\natseq{0}{t}}$. All nodes that can be reached 
from a node $i \in \nodestree{t \in \natseq{0}{\hor-1}}{\tree}$, are 
called child nodes of $i$, and are denoted by $\childtree{i}{\tree}$. 
Conversely, the ancestor node of a node $i \in \nodestree{\natseq{1}{\hor}}
{\tree}$ is denoted by $\anctree{i}{\tree}$. Using this notation, the 
optimization over policies $\pi$ can be reformulated as optimizing over 
a sequence of predicted states and input $(x_t, u_t)_{t=0}^{\hor-1}$, where 
a tuple $(x^{i}, u^{i})$ is assigned to each non-leaf node $i \in 
\nodestree{t \in \natseq{0}{\hor-1}}{\tree}$, and possible values for the 
terminal state $x^{l}$ to each leaf node $l \in \nodestree{\hor}{\tree}
$. We will use this representation of the problem to establish recursive 
feasibility in the next section.  

\subsection{Recursive feasibility of the nominal problem}\label{sec:recursive-feasibility}

In this section, we describe a simple procedure to obtain 
a robust control invariant set $\Xfeas_{\hor}$, and show that 
by imposing almost sure inclusion of the terminal state $x_\hor$ 
in this set, recursive feasibility of the nominal stochastic MPC 
problem can be established. In \Cref{sec:numerical-experiments}, we numerically compare the implications on the required safety distance with the prescriptions by the RSS framework described in~\cite{shalev-shwartz_formal_2017}.

\begin{defn}[Robust control invariant set]
    Let $\Xfeas$ denote a set of feasible states and $\Ufeas$ the set 
    of feasible control actions. A set $\rinv \subseteq \Xfeas$ is 
    called a \emph{robust control invariant} (RCI) set for the system~\eqref{eq:dynamics} if for all $x \in \rinv$, there exists a $u \in \Ufeas$ such that $f(x,u,w) \in \rinv, \forall w \in \W$.
\end{defn}

\begin{defn}[Maximal robust control invariant set] \sloppy
    An RCI set $\maxrinv$ is called the \emph{maximal robust control invariant} (MRCI) set, if for every other RCI set $\rinv$, it holds that $\rinv \subseteq \maxrinv$.
\end{defn}

\begin{defn}[Robust positively invariant set]
    For a given control law $\kappa: \Re^{\ns} \rightarrow \Re^{\na}$, a set $\rinv_\kappa \subseteq \Xfeas$ is a \emph{robust 
    positively invariant} (RPI) set for the system~\eqref{eq:dynamics} if 
    for all $x \in \rinv_\kappa$, it holds 
    that $\kappa(x) \in \Ufeas$ and $f(x, \kappa(x), w) \in \rinv_\kappa, \forall w \in \W$. Note that any RPI set is necessarily RCI. 
\end{defn}

For notational convenience, we construct a set $\Xs \subseteq \Xr \times \W$, akin to the~\emph{stochastic feasibility set} defined by~\cite{korda2011strongly}. It contains all augmented states $(x,w)$ that 
are feasible and for which a feasible input exists, with respect to both the soft constraints~\eqref{eq:nominal-MPC-crm-constraint} and hard constraints~\eqref{eq:nominal-MPC-state-input-constraint}: 
\begin{equation} \label{eq:stochastic-feasibility-set} 
    \Xs \dfn \left\{ (x,w) \middle|
        \begin{array}{c}
            x \in \Xr, w \in \W, \exists u \in \Ufeas : \\ \AVAR_\delta^{\row{\transmat}{w}}[g(f(x,u,w')\mid(x,w)] \leq 0\,\\
        w' \sim \row{\transmat}{w},
        \end{array}
        \right\}.
\end{equation}

Our goal is to compute a sufficiently large terminal constraint set $\Xfeas_\hor$, such that $\Xfeas_\hor \times \W \subseteq \Xs$. To this end, we first explicitly define a simple polyhedral RPI subset of $\Xr$ for the system~\eqref{eq:double-integrator-model} as shown in the following result. By iteratively expanding this set, we can then obtain an inner approximation of the MRCI set.

Let $\cmin \dfn \min_{w \in \W} c_w$ denote the parameter of the target vehicle model~\eqref{eq:TV-model} corresponding to the maximal deceleration, where we assume that 
$-\nicefrac{1}{\Ts} \leq \cmin < 0$.
\begin{prop}
Let us define the linear state feedback policy $u = Kx$, where 
$K \dfn \smallmat{0 & \cmin & 0}$, and the corresponding candidate RPI set
\begin{equation*} 
    \rinv_K \dfn \left \{ x \in \Re^{\ns} \middle| 
    \begin{array}{l}
        \frac{\amin}{\cmin} \geq x_2 \geq 0,\,
        x_3 \geq x_2, \\ 
        \vmax \geq x_2,\, 
        g(x) = -x_1 \leq 0
     \end{array}
    \right\}.
\end{equation*}
    The following statements hold:
    \begin{inlinelist}
        \item \label{statement-1} $\rinv_K$ is RPI for the dynamics~\eqref{eq:double-integrator-model} and policy $u = Kx$;
        \item \label{statement-2} $Kx \in \Ufeas$ for every $x \in \rinv_K$,
        with $\Ufeas$ as defined in~\eqref{robust-constraint-set-inputs}; and
        \item \label{statement-3} $\rinv_K \times \W \subseteq \Xs$
    \end{inlinelist}.
\end{prop}
\begin{pf}
    Statement~\ref{statement-1} is shown by applying the dynamics~\eqref{eq:double-integrator-model} to each of the defining constraints. Suppose that $x \in \rinv_K$ and let 
    $x^{+} \dfn f(x, Kx, w), w \in \W$ denote the uncertain successor state. By the assumption $-\nicefrac{1}{\Ts} \leq \cmin < 0$, we have that 
    \[
        x_2^+ = x_2 \smashoverbracket{(1 + \Ts \cmin)}{\in [0, 1)} \Rightarrow \max\left \{\tfrac{\amin}{\cmin}, \vmax \right \} \geq x_2^+ \geq 0.
    \]
    By definition of $\cmin$, and by the constraint 
    $x_3 \geq x_2$, we have for all $w \in \W$ that
    \[x_3^{+} \geq (1 +  \Ts \cmin) x_3 \geq (1 + \Ts \cmin) x_2 = x_2^+,
    \] 
    and $x_1^+ = x_1 + \Ts (x_3 - x_2) \geq 0$, thus $g(x^+) \leq 0$.  

    Statement~\ref{statement-2} follows by the assumption $-\nicefrac{1}{\Ts} \leq \cmin < 0$ combined with the first condition in the definition of $\rinv_K$: 
    \[
        \amax \geq 0 \geq \cmin x_2 \geq \amin.
    \]
    Finally, statement~\ref{statement-3} follows from the following two observations. $\rinv_K \subseteq \Xr$ and $x \in \rinv_K \Rightarrow g(x) \leq 0$. Indeed, by statement~\ref{statement-1}, $\exists u \in \Ufeas$, such that
        \[
            \begin{aligned}
                0 &\geq \max_{w' \in \W} \{g(f(x,u,w'))\} \\
                 &\geq \AVAR_\delta^{\row{\transmat}{w}}[g(f(x,u,w')\mid(x,w)],
            \end{aligned}
        \]
        for all $w, w' \in \W$.  
\end{pf}

We can now iteratively expand $\rinv_K$, to obtain the following iterates \cite[Alg. 2.1]{Kerrigan:2000}
\begin{equation*} 
    \rinv^{(i+1)} = \pre\left(\rinv^{(i)}\right) \cap \Xr \cap \Xc, \; \rinv^{(0)} = \rinv_K,
\end{equation*}
where $\pre(\rinv^{(i)}) \dfn \{x \in \Re^{\ns} | \exists u \in \Ufeas: f
(x, u, w) \in \rinv^{(i)}, \forall w \in \W\}$ denotes the pre-set of 
$\rinv^{(i)}$. Note that since all involved sets are polyhedral, the pre-set can be easily computed using standard techniques~(\cite{borrelli_predictive_2017}).
From~\cite[Prop. 2.6.1]{Kerrigan:2000},
it then follows that for all $i \in \N$, $\rinv^{(i)}$ is RCI. Therefore, we may choose to terminate after any finite number of 
iterations, and still retain guaranteed recursive feasibility, as we will now show.

\begin{defn}[Recursive feasibility]
    We say that an MPC controller is recursively feasible if the existence of a feasible solution $\pi_{|t} = (\kappa_{i})_{i \in \natseq{0}{\hor-1}}$ to the optimal control problem with initial state $(x, w) \in \Xs$ implies \textit{almost surely} that there exists a feasible solution to the optimal control problem with initial state $(f(x, \kappa_{0}(x, w), w'), w')$, $w' \sim \row{\transmat}{w}$.
\end{defn}
\begin{figure}
    \centering
    \begin{minipage}[t]{0.47\columnwidth}
        \centering
        \includegraphics{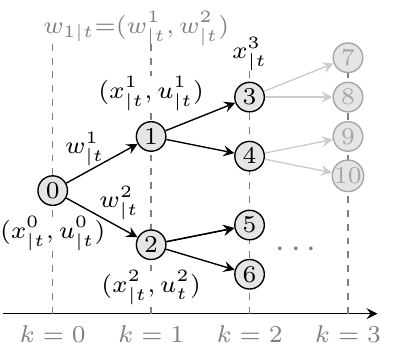}
        \scriptsize{$\tree_t$}
    \end{minipage} \hfill
    \begin{minipage}[t]{0.47\columnwidth}
        \centering
        \includegraphics{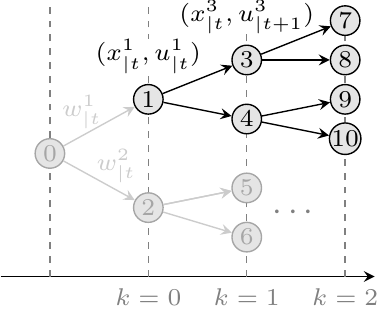}
        \scriptsize{$\tree_{t+1}$ for $w_{0|t+1} = w_{|t}^1$}
    \end{minipage}
    \caption{Illustration of the correspondence between the solutions of the optimal control problem over scenario trees $\tree_t$ and $\tree_{t+1}$ constructed at subsequent time steps, for a problem of horizon $\hor = 2$.}
    \label{fig:trees}
\end{figure}
\begin{thm}[Nominal recursive feasibility] \label{thm:recursive-feasibility}
If $\Xfeas_{\hor}$ is RCI and $\Xfeas_{\hor} \times \W \subseteq \Xs$, then the nominal stochastic MPC problem is recursively feasible. 
\end{thm}

\begin{pf}
    Suppose that at a given time $t$, a feasible solution exists, and
    let us denote the corresponding predicted states and sequences by 
    $(x_{k|t}, u_{k|t}, w_{k|t})_{k \in \natseq{0}{N-1}}$, $(x_{N|t}, w_{N|t})$. We represent these predictions on a scenario 
    tree $\tree_t$. Similarly, let us denote by $\tree_{t+1}$ the scenario tree spanned by the candidate predictions $(x_{k|t+1}, u_{k|t+1}, w_{k|t+1})_{k \in \natseq{0}{N-1}}$, $(x_{N|t+1}, w_{N|t+1})$ --- the feasibility of which is to be proven.

    $\tree_{t+1}$ is constructed by selecting the subtree of $\tree_t$ consisting of only the nodes with common ancestor $i \in \nodestree{1}{\tree_t}$ corresponding to the observed value of $w_{0|t+1}$, and extending it by one stage, as illustrated in~\Cref{fig:trees}. Thus,
    \begin{equation*}
        \nodestree{k \in \natseq{0}{\hor-1}}{\tree_{t+1}} \subset \nodestree{k \in \natseq{1}{\hor}}{\tree_{t}},
    \end{equation*} 
    and so, all non-leaf nodes of $\tree_{t+1}$ must have a corresponding node in $\tree_t$.
    \begin{proofsteps}
        \item \label{proof-recfeas-step-1} First, observe that all states and inputs stored in the non-leaf nodes of $\tree_t$ remain valid at time $t+1$. Indeed, 
        by definition, we have that for all $i \in \nodestree{k \in \natseq{1}{\hor-1}}{\tree_{t}}$, $(x_{|t}^i, w_{|t}^i, u_{|t}^i) \in \Xs \times \Ufeas$. Since the feasible sets $\Xs$ and $\Ufeas$ do not change from $t$ to $t+1$, these values remain feasible.
        \item \label{proof-recfeas-step-2} Furthermore, since for all predicted states $(x_{|t}^l, w_{|t}^l)$ in the leaf nodes $l \in \nodestree{\hor}{\tree_t}$ it holds that $(x_{|t}^l, w_{|t}^l) \in \Xfeas_\hor \times \W \subseteq \Xs$, these states remain feasible at the corresponding nodes at stage $\hor -1$ in $\tree_{t+1}$.   
        \item \label{proof-recfeas-step-3} It remains to verify that for all leaf nodes $l \in \nodestree{\hor}{\tree_t}$ at time $t$, a feasible input $u^l_{|t+1}$ exists,
        such that $f(x^l_{|t}, u^{l}_{|{t+1}}, w) \in \Xs$ for all $w \in \W: \elem{\transmat}{w^l_{t}}{w} > 0$. 
        Since $x_{|t}^l \in \Xfeas_\hor$ for all $l \in \nodestree{\hor}{\tree_t}$, the result follows immediately from the robust control invariance of $\Xfeas_\hor$.
    \end{proofsteps}
\end{pf}

\section{Distributionally robust formulation} \label{sec:dist-rob-mpc}

We now move to the more realistic setting in which the measure $\prob$, and by extension the transition matrix $\transmat \in \Re^{\dimProbSpace \times \dimProbSpace}$ governing the Markov chain is unknown.
In this setting, we need to resort to data-driven 
estimates of the transition probabilities, which are subject to some 
level of ambiguity. In order to cope with this ambiguity, while maintaining the established recursive feasibility, we adopt a distributionally robust approach, which leads to a modified version of the MPC 
problem~\eqref{eq:nominal-problem}.
\subsection{From Markovian data to ambiguity sets} \label{sec:data-driven-ambiguity}
Suppose we are given a sample $W = \{w_i\}_{i=1}^{\nsample}$ of $\nsample$ observations from a Markov chain with unknown transition matrix $\transmat$. To simplify matters, we partition $W$ into subsets $W_j \subseteq W$, 
$j \in \W$, which contain only the transitions 
that originated in mode $j$. That is, $W_j \dfn \{w_i \in \W \mid w_{i - 1} = j \}$.
Due to the Markov property, the samples $w \in W_j$ are independent and identically distributed (\iid{})  
with distribution $\row{\transmat}{j}$, i.e., the $j$th row of the
transition matrix. 
We compute the empirical distributions of $W_j$ to obtain estimates $\row{\hat{\transmat}}{j}$ for the transition probabilities. That is,
    \begin{equation} \label{eq:empirical-distribution}
        \begin{aligned}
            \elem{\hat{\transmat}}{j}{i} &\dfn
            \begin{cases}
                \tfrac{1}{\nsample_j} {\sum_{w \in W_j} 1_{w = i}}, & \text{if } \nsample_j > 0,\\
                \tfrac{1}{\dimProbSpace}, & \text{otherwise,}
            \end{cases}
        \end{aligned}
    \end{equation} 
for all $i, j\in \W$, where $\nsample_j \dfn |W_j|$ is the number of samples in each subset of the data. 
Given an arbitrary confidence level $\alpha \in (0,1]$, we can now for each such estimate $\row{\hat{\transmat}}{j}$, use the results in~\cite{schuurmans2019safe} to define an \emph{ambiguity set} 
\begin{equation} \label{eq:def-amb-tv}
    \ambTV_{r_j}(\row{\hat{\transmat}}{j}) \dfn \{ p \in \simplex_{\dimProbSpace} \mid \| p - \row{\hat{\transmat}}{j} \|_1 \leq r_j\},
\end{equation}
where the radius
\begin{multline} \label{eq:radius-McDiarmid}
    r_j = r(\alpha, \dimProbSpace, \nsample_j) \\\dfn \sqrt{-\frac{2 \ln(\alpha)}{\nsample_j}} + \sqrt{\frac{2 (\dimProbSpace-1)}{\pi \nsample_j}} + \frac{4 \dimProbSpace^{\nicefrac{1}{2}} (\dimProbSpace-1)^{\nicefrac{1}{4}}}{\nsample_j^{\nicefrac{3}{4}}}, 
\end{multline}
is computed such that
\(
    \prob[\row{\transmat}{j} \in \ambTV_{r_j}(\row{\hat{\transmat}}{j})] \geq 1 - \alpha.
\)

By the dual risk representation~\eqref{eq:risk-dual}, the computed ambiguity sets $\ambTV_{r_j}(\row{\hat{\transmat}}{j})$ implicitly define coherent risk measures.
Thus, by replacing the now unknown probability distributions in the formulation of the nominal MPC problem~\eqref{eq:nominal-problem} by the worst-case distribution in the estimated ambiguity sets, we transform it to a \emph{risk-averse} MPC problem (\cite{sopasakis2019risk}), in which the ambiguity in the estimated transition matrices is accounted for. 

By collecting additional data samples during closed-loop operation 
-- that is, by increasing $n_j$ and therefore decreasing $r_j$ corresponding to mode $j$ -- the estimated transition probabilities $\row{\hat{\transmat}}{j}$ will converge to their true underlying values $\row{\transmat}{j}$ while the related ambiguity sets asymptotically shrink to the singletons
$\{\row{\transmat}{j}\}$.
As such, conservatism of the controller is gradually reduced throughout
closed-loop operation, while constraint satisfaction with respect to the true distributions is guaranteed with arbitrarily high probability.
The overall control scheme described in the next section, including online/offline learning, is summarized in~\Cref{alg:algorithm-dr-MPC}.  
\subsection{Risk-averse MPC formulation} \label{sec:risk-averse-mpc}
\subsubsection{Cost function}

The proposed distributionally robust approach
replaces the conditional expectations by conditional risk mappings based on the risk measures induced by the ambiguity sets~\eqref{eq:def-amb-tv}. 
For ease of notation we will for a given sequence of ambiguity sets $\bar{\amb} \dfn (\amb_j)_{j\in \W}$, denote the conditional risk mapping of the random stage costs as
\begin{align*}
    \crm{t}^{\bar{\amb}}[\ell(x_{t+1}, u_{t+1})] \dfn \max_{p \in \amb_{w_t}}\E_{\delta}^{p}[ \ell(x_{t+1}, u_{t+1}) \mid x_t, w_t].
\end{align*}

\subsubsection{Ambiguous chance constraints}
\label{sec:chance-constraints}
Since the implication~\eqref{eq:risk-constraint-implication} holds only
with respect to the true but unknown probability measure $\prob$, the risk
constraint \eqref{eq:nominal-MPC-crm-constraint} no longer guarantees
satisfaction of the original chance constraints in the current setting. We
will therefore impose it robustly with respect to all distributions in the
data-driven ambiguity sets $\ambTV_{r_j}(\row{\hat{\transmat}}{j})$, 
leading to the following definition.

\begin{defn}[Distributionally robust $\AVAR$] \sloppy
    Given a random vector $z \in \Re^n$ and an ambiguity set $\amb \subseteq \simplex_n$, we define the distributionally robust average value-at-risk of $z$ as   
    \begin{equation} \label{eq:robust-avar}
        \TVAVAR_{\delta}^{\amb}[z] \dfn \max_{\nu \in \amb} \AVAR_{\delta}^{\nu}[z].
    \end{equation}
\end{defn}
For the $\ell_1$-ambiguity set $\amb = \ambTV_{r}(\hat{p})$ of radius $r$ around an empirical estimate $\hat{p}$, we can use the definitions \eqref{eq:def-amb-tv} and \eqref{eq:def-AVAR} of $\ambTV_{r}(\hat{p})$ and 
$\amb_{\AVAR_{\delta}^{\nu}}$ to express~\eqref{eq:ambiguity-avar} explicitly as 
\begin{equation*} 
\TVAVAR_{\delta}^{\ambTV_{r}(\hat{p})}[z] = \max_{\pi, \nu \in \simplex_n} \left \{\trans{\pi} z \middle| \begin{array}{c}
    \| \nu - \hat{p} \|_1 \leq r,    \\ 
    \pi \leq \nicefrac{\nu}{\delta}
    \end{array}  
     \right\}. 
\end{equation*}

Recall that we assume that the radius $r$ in the definition of the ambiguity set is chosen to satisfy
\[
\prob[p \in \ambTV_{r}(\hat{p})] \geq 1 - \alpha. 
\]
Therefore we have that with probability at least $(1-\alpha)$, 
\(
    \AVAR_{\delta}^{p}[z] \leq \TVAVAR_{\delta,r}^{\ambTV_{r}(\hat{p})}[z],
\)
so that a constraint on a random value $z$ of the form 
\[
    \TVAVAR_{\delta,r}^{{\ambTV_{r}(\hat{p})}}[z] \leq 0,
\]
implies that 
\(
    \prob[z \leq 0] \geq 1-\epsilon,    
\)
where 
\[1-\epsilon \geq (1-\delta)(1-\alpha).
\] 
Thus, by replacing the $\AVAR$ risk measure used in the conditional 
risk constraints~\eqref{eq:nominal-MPC-crm-constraint} by $\TVAVAR$, 
satisfaction of chance constraint can still be guaranteed despite the incomplete knowledge of the transition matrix.

We summarize the modifications to the nominal problem formulation in the following definition.

\begin{defn}[Risk-averse MPC problem] \label{def:dr-mpc}
    \sloppy For a given initial state $x \in \Re^{\ns}, w \in \W$,
    and sequence of ambiguity sets $\ambSeq \dfn (\amb_j \subseteq \simplex_\dimProbSpace)_{j\in \W}$, 
    the risk-averse OCP comprises of computing an 
    $\hor$-step sequence of admissible policies $\pi = (\kappa_i)_{i \in \natseq{0}{\hor-1}}$, with 
    $\kappa_k : \Re^{\ns} \times \W \rightarrow \Re^{\na}$ that 
    solve the optimization problem 
        \begin{subequations}\label{eq:dr-problem}
            \begin{align}
            \minimize_{u_0}\;& \ell(x_0, u_0) + \inf_{u_1} \crm{0}^{\bar{\amb}} \Big[ \ell(x_1, u_1)\notag {}+{}\ldots \\
            + \inf_{u_{\hor-1}}& \crm{\hor-2}^{\bar{\amb}} \big[ \ell(x_{\hor-1}, u_{\hor-1}){}+{} \crm{\hor-1}^{\bar{\amb}} \big[ \ell_{\hor}(x_\hor) \big] \cdots
            \Big]
            \bigg] 
            \end{align}
            subject to
                \begin{align}
                    &x_0= x, w_0 = w, \\
                    &x_{k+1} = f(x_k, u_k, w_{k+1}),\, \forall k \in \natseq{0}{\hor-1},\\
                    \label{eq:dr-MPC-state-input-constraint}
                    &u_k = \kappa_k(x_k, w_k) \in \Ufeas, x_k \in \Xr,\,
                    \begin{array}{l}
                        \forall k \in \natseq{0}{\hor-1},\\
                        \forall \w_k \in \W^k,
                    \end{array}\\
                    \label{eq:dr-MPC-crm-constraint}
                    &\TVAVAR_{\delta}^{\amb_{w_{k}}}[g(x_{k+1}) \mid x_k, w_k] \leq 0,\, \begin{array}{l}
                    \forall k \in \natseq{0}{\hor-1},\\
                    \forall \w_k \in \W^{k}, \end{array}\\ \label{eq:terminal}
                    &x_\hor \in \Xfeas_\hor, \, \forall \w_{\hor} \in \W^\hor,
                \end{align}
            \end{subequations}
            where we introduced the shorthand $\w_k \dfn (w_i)_{i=1}^{k}$.
            The corresponding learning-based MPC scheme is presented in~\Cref{alg:algorithm-dr-MPC}.
\end{defn}
\begin{rem}
    Without knowledge of the true distributions, imposing constraints \textit{almost surely} -- even for all distributions in the ambiguity set -- is no longer sufficient to guarantee recursive feasibility, since with a probability of at most $\alpha > 0$, a nonzero transition probability to a given mode is not reflected in any probability vector in the used ambiguity set. As a result, a feasible solution at a given time cannot be used to guarantee the existence of a feasible solution in the next. Therefore, we impose constraints at stage $k$ \emph{for all} realizations of $\w_k$.     
\end{rem}

As mentioned earlier, this problem can be stated as
a finite-dimensional optimization problem using a scenario tree representation.
Furthermore, \cite{sopasakis2019riskC} show that the risk-averse OCP~\eqref{eq:dr-problem} can be 
reformulated tractably, provided that the involved risk measures
are conic representable. We say that a risk measure $\rho: \Re^n \rightarrow \Re$ is conic representable if 
\begin{equation*}
    \rho[z] = \max_{\mu \in \Re^n, \nu \in \Re^r} \{\trans{\mu} z \mid E\mu + F \nu \preceq_{\cone} b\},
\end{equation*}
for some matrices $E, F$ and a vector $b$ of appropriate dimensions and a closed, convex cone $\cone$. It is straightforward to verify that for the risk measures involved, namely $\TVAVAR$ and the risk measure induced by the $\ell_1$-ambiguity set~\eqref{eq:def-amb-tv}, consisting only of linear (in)equalities, this is indeed the case.
Moreover, since the model described in~\Cref{sec:modeling} has quadratic costs and linear constraints, the reformulation of the problem in~\cite{sopasakis2019riskC} leads to a convex, quadratically constrained, quadratic program (QCQP), which we can solve efficiently using off-the-shelf solvers.
\subsection{Recursive feasibility of the risk-averse MPC problem}
Finally, we show recursive feasibility of the learning-based MPC scheme by 
slightly adapting the proof of~\Cref{thm:recursive-feasibility}.

Let us for a given sequence  $\bar{\amb} = (\amb_j)_{j\in\W}$ of ambiguity sets define a set $\hat{\Xs}(\bar{\amb})$, analogously to $\Xs$ in the nominal case: 
\begin{equation} \label{eq:dr-feasibility-set} 
    \hat{\Xs}(\bar{\amb}){\dfn}\left\{ (x,w) \middle|
    \begin{array}{c}
        x \in \Xr, w \in \W, \exists u \in \Ufeas : \\ \TVAVAR_\delta^{\amb_w}[g(f(x,u,w'){\mid}(x,w)]{\leq}0\,\\
    w' \sim \row{\transmat}{w},
    \end{array}
    \right\}.
\end{equation}
\begin{thm}[Risk-averse recursive feasibility] \label{thm:dr-recursive-feasibility}
    \sloppy
    If for all time steps $t$ and $t+1$, the risk-averse MPC problem~\eqref{eq:dr-problem} is instantiated with ambiguity sets $\ambSeq_t = (\amb_{t,j})_{j\in \W}$ and $\ambSeq_{t+1} = (\amb_{t+1,j})_{j \in \W}$, such that 
    \begin{equation} \label{eq:nested-ambiguity} 
        \amb_{t+1,j} \subseteq \amb_{t,j},\, \forall j \in \W,
    \end{equation}
    then, the learning risk-averse MPC scheme is recursively feasible.
\end{thm}
\begin{pf}
    The proof is along the lines of that of~\Cref{thm:recursive-feasibility}, given the following modifications.
    Since in the current setting, the ambiguity sets may change between subsequent instances of the OCP, so may the stochastic feasible set. 
    Thus, for step~\ref{proof-recfeas-step-1} to hold, the following implication is required for all $i \in \nodestree{k \in \natseq{1}{\hor-1}}{\tree_{t}}$:
    \[
         (x_{|t}^i, w_{|t}^{i}) \in \hat{\Xs}(\bar{\amb}_{t}) \Rightarrow 
         (x_{|t}^i, w_{|t}^{i}) \in \hat{\Xs}(\bar{\amb}_{t+1}),
    \]
    or equivalently, $\hat{\Xs}(\bar{\amb}_{t}) \subseteq \hat{\Xs}(\bar{\amb}_{t+1})$.
    This, in turn, follows from the condition~\eqref{eq:nested-ambiguity}
    by filling in the expression~\eqref{eq:robust-avar} for the $\TVAVAR$ risk measure in the definition~\eqref{eq:dr-feasibility-set} of the feasible set.
    Step~\ref{proof-recfeas-step-2} requires that $\Xfeas_\hor \times \W \subseteq \hat{\Xs}(\bar{\amb}_{t}) \Rightarrow \Xfeas_\hor \times \W \subseteq \hat{\Xs}(\bar{\amb}_{t+1})$, which follows from the same argument.
    Step~\ref{proof-recfeas-step-3} relies solely on the robust control invariance of the terminal constraint set and thus remains valid.
\end{pf} 
\begin{rem}
    Note that for the nominal stochastic approach, no ambiguity is taken into account, i.e., $\amb_j = \{\hat{\row{\transmat}{j}}\},\, \forall j \in \W$. Therefore, the nested ambiguity condition~\eqref{eq:nested-ambiguity} can only be satisfied if the transition probabilities are estimated once and kept fixed afterwards. 
\end{rem}
\begin{algorithm}
    \caption{Learning risk-averse MPC}
    \label{alg:algorithm-dr-MPC}
    \small
    \begin{algorithmic}
        \Require $x_0, w_0$, $W$
        \For {$j \in \natseq{1}{\dimProbSpace}$} \Comment{Optional offline learning step}
            \State Initialize $\row{\hat{\transmat}}{j}$, $r_j$ using \eqref{eq:empirical-distribution}--\eqref{eq:radius-McDiarmid}
            \State $\ambSeq_j \gets \ambTV_{r_j}(\row{\hat{\transmat}}{j})$
        \EndFor
        \For {$k \in \N_{0}$} \Comment{Learning MPC} 
            \State $(\kappa_i)_{i=0}^{\hor-1} \gets$ solve~\eqref{eq:dr-problem} given $x_{k}$, $w_{k}$, $\ambSeq$ 
            \State $(x_{k+1},w_{k+1}) \gets$ Apply $u = \kappa_0(x_k,w_k)$ to system~\eqref{eq:double-integrator-model} and observe state
            \State $W \gets W \cup \{w_{k+1}\}$
            \State $j \gets w_k$
            \State Update $\row{\hat{\transmat}}{j}$, $r_j$ using \eqref{eq:empirical-distribution}--\eqref{eq:radius-McDiarmid}
            \If {$\ambTV_{r_j}(\row{\hat{\transmat}}{j}) \subset \amb_j$} \Comment{Update ambiguity if~\eqref{eq:nested-ambiguity} is satisfied}
                \State $\amb_j \gets \ambTV_{r_j}(\row{\hat{\transmat}}{j})$
            \EndIf
        \EndFor
    \end{algorithmic}
\end{algorithm}
\section{Numerical simulations} \label{sec:numerical-experiments} 

\subsection{Terminal constraint sets}
For the considered set-up, the RSS model described in~\cite{shalev-shwartz_formal_2017} derives a 
minimal safety distance required for guaranteed collision avoidance. 
It involves computing the distances $\de(x_2), \dt(x_3)$ required for the ego vehicle and target vehicle respectively to come to a halt in an emergency braking scenario, as a function of their initial velocities $x_2, x_3$. The minimal required distance is given by $h_{\submin, \mathrm{RSS}}(x_2,x_3) \dfn [\de(x_2) - \dt(x_3)]_+$. Although derived for continuous-time systems, the derivation can be easily repeated for the discrete-time model at hand.
Somewhat surprisingly, however, the $[\argdot]_+$-operator involved in the definition of $h_{\submin, \mathrm{RSS}}(x_2,x_3)$ prohibits the set $\Xrss \dfn \{x \mid x_1 \geq h_{\submin, \mathrm{RSS}}(x_2,x_3)\}$ from 
being RCI, unless specific conditions on the system parameters are met.

Similarly, for a given pair of velocities $x_2$ and $x_3$, the iteratively computed terminal constraint sets $\rinv^{(i)}$ can be 
associated to a minimal safety distance $h_{\submin}^{(i)}(x_2,x_3) \dfn \min \{h \mid \trans{\smallmat{h & x_2 & x_3}} \in \rinv^{(i)} \}$, where we set $\h_{\submin}^{(i)} = \infty$ if no feasible solution exists. 

\begin{figure}[htb!]
    \centering
    \includegraphics{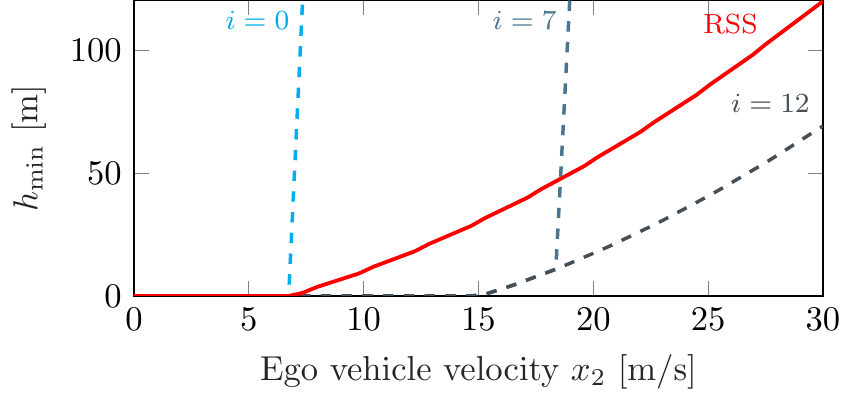}
    \caption{Minimal safety distances $h^{(i)}_{\submin}$ and $h_{\submin, \mathrm{RSS}}$, for $\vmax=40\si[per-mode=symbol]{\meter\per\second}$,
    $\amin = -5 \si[per-mode=symbol]{\meter\per\second\squared}$,
    $\cmin = -0.33 \si{\per\second}$ and a fixed target vehicle velocity $x_3 = 20\si[per-mode=symbol]{\meter\per\second}$.}
    \label{fig:distances}
\end{figure}

\Cref{fig:distances} shows the safety distance according to 
both approaches as a function of $x_2$. Note that the initial set $\rinv^{(0)}$ is more conservative than RSS. However, after $i=12$ iterations, $\rinv^{(i)}$ has converged and yields a smaller safety distance than RSS for all values of $x_2$. Thus, we find that in practice, the requirement of the terminal set to be RCI introduces no conservatism over the hand-crafted safety distance 
provided by RSS.

\subsection{Closed-loop simulations} \label{sec:experiments-exact}

The following experiments demonstrate the benefit of the proposed 
learning-based MPC scheme in \Cref{alg:algorithm-dr-MPC} (referred to as the \emph{risk-averse} approach), as compared to 
the two extreme variants obtained by taking $\amb_j = \{\row{\hat
{\transmat}}{j}\}$ and $\amb_j = \simplex_\dimProbSpace$, for all $j \in 
\W$. We refer to these as the \emph{stochastic} and \emph{robust} approach,
respectively.
For the stochastic approach, we set the tolerated chance constraint violation probability to $\delta_{\textrm{s}} = 0.1$, and for the risk-averse controller, we choose $\alpha = \delta = 0.05$, such that
$(1-\alpha)(1-\delta) \approx 1-\delta_\textrm{s}$.
All used controller settings are as summarized in \Cref{tab:parameters-exact}, unless otherwise specified. The (unknown) transition matrices used in the experiments are  
\[
    \transmat_{p} = \smallmat{0.92&    0.04 &   0.02 &   0.02\\
                          0.29&    0.50 &   0.09 &   0.12\\
                          0.26&    0.21 &   0.36 &   0.17\\
                          0.31 &   0.25 &   0.23 &   0.21} \text{ and }
    \transmat_{s} = \smallmat{0.29 &  0.7 &  0.009 & 0.001 \\
                        0.09 & 0.90 & 0.009& 0.001\\
                        0.4 & 0.29 & 0.3 & 0.01\\
                        0.048 & 0.001 & 0.001 & 0.95}.
\]
The optimal control problems are formulated using Yalmip (\cite{Lofberg2004}) and solved using MOSEK (\cite{mosek}) on an Intel Core i7-7700K CPU.

\begin{table}[ht!]
    \footnotesize
    \centering
    \caption{Default controller settings.}
    \label{tab:parameters-exact}
    \begin{tabular}{@{}ccccc@{}}
    \toprule
    $(q,r)$ &
    \begin{tabular}[c]{@{}c@{}}
    $\Ts$
    $[\si{\second}]$
    \end{tabular}
    & $\hor$ & 
    \begin{tabular}[c]{@{}c@{}}
        $(\vref, \vmax)$
        $[\si[per-mode=symbol]{\meter\per\second}]$
    \end{tabular}&
    \begin{tabular}[c]{@{}c@{}}
        $(\amin, \amax)$
        $[\si[per-mode=symbol]{\meter\per\second\squared}]$
    \end{tabular}\\
    \midrule
    $(5,10)$ & $0.5$ & $3$ & $(30, 40)$ & $({-4}, 5)$ \\
    \bottomrule
\end{tabular}
\end{table}

\subsubsection{Performance} For a fixed initial state,
we performed 100 randomized simulations of 50 time steps for the three controllers with prediction horizon $\hor=5$.  
The target vehicle parameters are $
(c_i)_{i\in \W} = \smallmat{1.13& {-0.02}& {-0.33} & {-0.16}}$ and the true transition matrix is set to $P=P_p$. The average solver time for these experiments was $0.45 \si{\second}$.

We compare the performance of the controllers by computing the closed-loop 
cost over each realization. We conducted this experiment both with and 
without offline learning. In the former case, all transition probabilities 
are estimated online, whereas in the latter, a sequence of 5000 draws from the Markov chain is provided to the controller before deployment. 

\begin{figure}[htb!]
    \centering
    \includegraphics{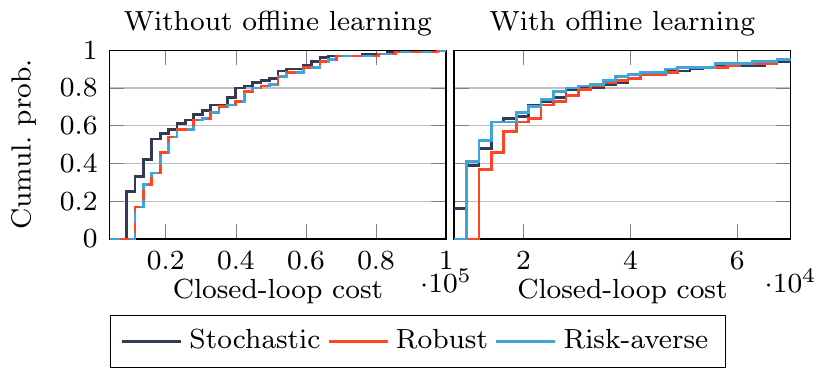}
    \vspace{-3mm}
    \caption{Empirical cumulative distribution of the closed-loop cost over
    100 randomized simulations.}
    \label{fig:cumulative-costs-true-model}
\end{figure}

\Cref{fig:cumulative-costs-true-model} shows the empirical cumulative 
distribution of the closed-loop costs with and without offline learning. 
We observe that due to the initial lack of data, the risk-averse 
controller selects a large ambiguity set, which renders its behavior 
indistinguishable from that of the robust controller. The stochastic 
approach, on the other hand, introduces no such conservatism and thus 
achieves lower costs more frequently than the competing controllers. 
As the risk-averse controller observes more data (\Cref{fig:cumulative-costs-true-model}, right), its conservatism decreases, allowing it to achieve a cost distribution that 
closely resembles that of the stochastic approach, while still providing the same recursive feasibility guarantees as the robust approach.

\subsubsection{Safety} \label{sec:safety}
In the following experiment, we use the target vehicle parameters
$(c_i)_{i\in \W} = \smallmat{1.1 & 0 & {-0.5} & {-1}}$ and 
transition matrix $P = \transmat_{s}$. In order to simulate a low-probability emergency situation, we force 
the Markov chain to switch to mode 4 at a single fixed time step during each simulation, which corresponds to a harsh braking maneuver of the target vehicle. Note that from any mode $i \in \W$, there is a nonzero switching probability to mode 4. Therefore, the simulated trajectories correspond to possible realizations for which infeasibility 
of the OCP is not acceptable.

We repeated this simulation for 100 realizations of 200 steps, and with increasing sample sizes $n$ for offline learning. The average solver time for this experiment was $0.036\si{\second}$.

\begin{figure}[htb!]
\centering
\includegraphics{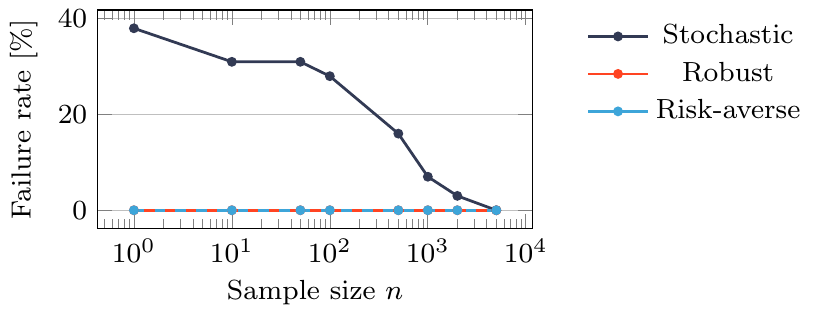}
\vspace{-3mm}
\caption{Percentage of infeasible realizations for the emergency braking scenario (out of 100 realizations).}
\label{fig:failures_stochastic}
\end{figure}

\Cref{fig:failures_stochastic} shows that with minimal offline learning, 
the stochastic controller fails to find a feasible solution in 
$38\%$ of realizations. As $\nsample$ increases and estimated 
distributions become more accurate, this fraction decreases, yet it requires a sample size $\nsample=5000$ to reduce the number of infeasible realizations to zero for this particular experiment. On the contrary, \Cref{thm:dr-recursive-feasibility} guarantees recursive feasibility for the risk-averse and the robust approach regardless of $\nsample$, as confirmed by the experiment.   

\section{Conclusion}

We proposed a learning-based risk-averse approach 
towards MPC for ACC applications with Markovian driver models. This 
framework allows us to utilize collected data to improve performance of 
the controller with respect to the robust approach, while retaining safety 
guarantees through provable recursive feasibility. These benefits were 
illustrated by means of closed-loop simulations. 
In future work, we plan to perform more extensive experiments using real-world driving data as well as for more general automated driving set-ups involving collision avoidance.

\bibliography{references}         

\end{document}